%% file: DTF.tex
\documentclass[
 	screen,
 	nonacm,
]{acmart}

\input{preamble}

\begin{document}

	\title{Discrete-event Tensor Factorization: Learning a Smooth Embedding for Continuous Domains}
	
	\author{Joey De Pauw}
	\email{joey.depauw@uantwerpen.be}
	\orcid{0000-0002-1417-922X}
	\affiliation{%
	  \institution{University of Antwerp}
	  \streetaddress{Middelheimlaan 1}
	  \city{Antwerp}
	  \country{Belgium}
	}
	
	\author{Bart Goethals}
	\email{bart.goethals@uantwerpen.be}
	\orcid{0000-0001-9327-9554}
	\affiliation{%
		\institution{University of Antwerp}
		\streetaddress{Middelheimlaan 1}
		\city{Antwerp}
		\country{Belgium}
	}
	\affiliation{%
		\institution{Monash University}
		\city{Melbourne}
		\country{Australia}
	}

	\begin{abstract}
		\input{00_abstract.tex}
	\end{abstract}



	\maketitle
	
	\input{01_introduction.tex}
	\input{03_model.tex}

	\input{05_experiments.tex}
	\input{98_related_work.tex}

	\input{99_conclusions.tex}

	\begin{acks}
		This work was supported by the \grantsponsor{fwo}{Research Foundation --- Flanders (FWO)}{https://www.fwo.be/} [\grantnum{fwo}{11E5921N} to J. De Pauw] and the 
		\grantsponsor{gov}{Flemish Government}{https://airesearchflanders.be/} under the \grantnum[https://airesearchflanders.be/]{gov}{``Onderzoeksprogramma Artificiële Intelligentie (AI) Vlaanderen''} programme.
		The computational resources and services used in this work were provided by the HPC core facility CalcUA of the Universiteit Antwerpen, and VSC (Flemish Supercomputer Center), funded by the Research Foundation - Flanders (FWO) and the Flemish Government.
	\end{acks}

	\bibliographystyle{ACM-Reference-Format}
	\bibliography{references}

\end{document}

%% file: preamble.tex
\usepackage{hyperref}
\usepackage{amsmath}
\usepackage{graphicx}
\usepackage[american]{babel}
\usepackage{xcolor}
\usepackage{tabularx}
\usepackage{multirow}
\usepackage{siunitx}
\usepackage{tikz}
\usepackage{pgfplots}
\usepackage{pgfplotstable}
\usepackage{xfrac}
\usepackage[inline]{enumitem}

\pgfplotsset{compat=newest}
\usetikzlibrary{
    pgfplots.dateplot,
}

\usepgfplotslibrary{external} 
\pgfplotsset{
    every axis post/.style={
        cycle list name=exotic,
        /pgf/number format/.cd,
        1000 sep={},
    },
    axis lines = left,
    legend cell align={left},
}

\definecolor{P1}{HTML}{D81B60}
\definecolor{P2}{HTML}{1E88E5}
\definecolor{P3}{HTML}{FFC107}
\definecolor{P4}{HTML}{004D40}
\definecolor{P5}{HTML}{5D3A00}

\pgfplotscreateplotcyclelist{exotic2}{
    P1,every mark/.append style={fill=P1!80!black},mark=*\\
    P2,every mark/.append style={fill=P2!80!black},mark=square*\\
    P3,every mark/.append style={fill=P3!80!black},mark=pentagon*\\
    P4,every mark/.append style={fill=P4!80!white},mark=triangle*\\
    P5,every mark/.append style={fill=P5!80!white},mark=diamond*\\
}

\DeclareMathOperator*{\diag}{diag}

\newcommand*{\normsq}[1]{\left\lVert#1\right\rVert^2}
\newcommand*{\Xpred}{\ensuremath{\hat{X}}}
\newcommand*{\frobnormsq}[1]{\normsq{#1}_F}

\newcommand*{\diff}{\mathop{}\!\mathrm{d}}
\newcommand*{\R}{\mathbb{R}}
\newcommand*{\Bl}{\ensuremath{{B^{(l)}}}}
\newcommand*{\Cl}{\ensuremath{{C^{(l)}}}}
\newcommand*{\loss}{\ensuremath{\mathcal{L}}}
\newcommand*{\lossl}[1]{\ensuremath{\loss^{\textit{(#1)}}}}

\newcommand*{\citeauthorplain}[1]{\begin{NoHyper}\citeauthor{#1}\end{NoHyper}}

\definecolor{C1}{HTML}{EA8D2A}
\definecolor{C2}{HTML}{33673B}
\definecolor{C3}{HTML}{63CCCA}
\definecolor{C4}{HTML}{BCA3AC}
\definecolor{C5}{HTML}{016FB9}
\definecolor{C6}{HTML}{49416D}
\definecolor{C7}{HTML}{BB0A21}

\definecolor{iTALS}{named}{C2}
\definecolor{iTALS TA}{named}{iTALS}
\definecolor{iTALS TA2}{named}{iTALS}
\definecolor{iTALS TA4}{named}{iTALS}

\definecolor{iTALSx}{named}{C3}
\definecolor{iTALSx TA}{named}{iTALSx}
\definecolor{iTALSx TA2}{named}{iTALSx}

\definecolor{Popularity}{named}{C4}
\definecolor{Popularity2}{named}{Popularity}

\definecolor{DTF}{named}{C1}
\definecolor{DTF2}{named}{DTF}
\definecolor{DTFKernel}{named}{DTF}
\definecolor{DTF4}{named}{DTF}

\definecolor{EASE}{named}{C5}
\definecolor{EASE TA}{named}{EASE}
\definecolor{EASE Time}{named}{EASE}

\definecolor{WMF}{named}{C6}
\definecolor{WMF TA}{named}{WMF}
\definecolor{WMF TA2}{named}{WMF}

\definecolor{DMF}{named}{C7}

%% file: 00_abstract.tex

Recommender systems learn from past user behavior to predict future user preferences.
Intuitively, it has been established that the most recent interactions are more indicative of future preferences than older interactions.
Many recommendation algorithms use this notion to either drop older interactions or to assign them a lower weight, so the model can focus on the more informative, recent information.
However, very few approaches model the flow of time explicitly.

This paper analyzes how time can be encoded in factorization-style recommendation models. By including absolute time as a feature, our models can
learn varying user preferences and changing item perception over time. In addition to simple binning approaches, we also propose a novel, fully continuous time encoding mechanism.
Through the use of a polynomial fit inside the loss function, our models completely avoid the need for discretization, and they are able to capture the time dimension in arbitrary resolution.

We perform a comparative study on three real-world datasets that span multiple years, where long user histories are present, and items stay relevant for a longer time.
Empirical results show that, by explicitly modeling time, our models are very effective at capturing temporal signals, such as varying item popularities over time.
Despite this however, our experiments also indicate that a simple post-hoc popularity adjustment is often sufficient to achieve the best performance on the unseen test set. This teaches us that, for the recommendation task, predicting the future is more important than capturing past trends. As such, we argue that specialized mechanisms are needed for extrapolation to future data.

%% file: 01_introduction.tex

\section{Introduction}
\label{sec:introduction}

Dealing with time is a crucial aspect of recommender systems~\cite{rabiu2020recommender, rana2015study}. A user's taste can change as they grow older or pick up new interests~\cite{lo2018temporalmf}. Similarly, an item's popularity may vary over time as it grows outdated or the majority of users have seen it~\cite{sun2017dynamic}. Furthermore, in the context of collaborative filtering, the time when a user interacts with an item provides valuable information about the user's preferences and the item's relevance. For example, watching a blockbuster movie when it is released may indicate a different interest than watching it years later~\cite{koren2009collaborative}.

Incorporating this ever-changing nature of the recommendation task is far from trivial. With the end-goal of retrieving the most relevant content at the current time in mind, most approaches take a \emph{passive} approach to including time~\cite{rabiu2020recommender}. Among such techniques are:
\begin{enumerate*}[label=\arabic*)]
\item forgetting mechanisms~\cite{matuszyk2015forgetting}, that incrementally update existing models to adapt them to current trends.
\item weighting~\cite{ding2005time}, where a higher contribution is given to recent data, and
\item window-based approaches~\cite{steck2019collaborative, verachtert2022we}, where dynamic cutoffs ensure that only timely information is used.
\end{enumerate*}

However, with datasets spanning multiple years and long user histories being available (see Figure~\ref{fig:session_spans}), we argue for an \emph{active} approach to modeling time. Indeed, where the previous approaches all adapt the model to the current trends, they fail to capture the intricacy of when interactions occur and precisely how this influences what can be learned from them. The blockbuster example above shows the value of an absolute notion of time, that can capture popularity trends. Furthermore, relative time between interactions can also be considered, such as when two items are consumed close together, this implies a stronger similarity between the two~\cite{garg2019sequence, choi2021session}.

In this paper, we study different ways to encode \emph{absolute} time in factorization-style recommendation models. By explicitly modeling time as a factor, we can represent the flow of time in the learned weights, as demonstrated in Figure~\ref{fig:predictions}. With time being one of the most important contexts of an interaction~\cite{pagano2016contextual}, we demonstrate that models that account for temporal dynamics, can better capture some intricate dependencies and improve top-k recommendations.

Consider the classic user-item interaction matrix $X \in \{0, 1\}^{m \times n}$ with $m$ users and $n$ items. $X_{ui} = 1$ denotes an interaction of user $u$ with item $i$. Matrix factorization algorithms decompose this large and sparse matrix into the product of low-rank user and item matrices $P \in \R^{m \times k}$ and $Q \in \R^{n \times k}$~\cite{hu2008collaborative, pan2008one}. They represent the latent factors or embeddings of the users and items respectively. Due to the model being an approximation of rank $k$, it can predict missing entries in the matrix, which correspond to the items that a user has not yet interacted with.

To incorporate time as a feature, most approaches first discretize the time axis with a binning strategy~\cite{hidasi2012fast, hidasi2014factorization, hidasi2014approximate, depauw2025wtf}. This way, a discrete 3D-tensor $X \in \R^{m \times n \times l}$ can be constructed, where $l$ is the number of bins. For example, all interactions can be grouped into bins that span one day, to effectively get an interaction matrix per day, all stacked together into the tensor. From this input format, tensor decomposition methods can be used to learn the user and item embeddings, as well as time factors, one for each bin.

The binning however is a crude approximation and it leads to some drawbacks and assumptions. First, having too many bins results in more sparse data, because there are fewer interactions per bin. Secondly, the bins are typically independent of each other, meaning the arbitrary lengths and cutoff points can have a large and inconsistent impact on predictions. Thirdly, the computational cost and model size often depend on the amount of bins. Hence binning imposes a trade-off between, supporting a fine granularity by taking small intervals, and preserving enough data in each bin to be able to learn the model, by taking large enough intervals~\cite{koren2009collaborative}.

Our decomposition approach however, works directly on the continuous time axis and completely bypasses the discretization step and all of its drawbacks. The key insight is that the learned time factors are not independent of each other~\cite{pagano2016contextual, hidasi2014approximate}. Indeed, because user tastes and item perception change smoothly over time, we can assume that time factors close together should be similar. This allows us to introduce a polynomial fit of the time factors, directly into the decomposition. With this fit we effectively learn continuous weights for the time factor and circumvent the sparsity issue, as the coefficients of the fit are learned based on \emph{all} timestamps, and no binning is needed.

Section~\ref{sec:models} describes the theoretical background and our contributions in detail. In Section~\ref{sec:experiments} we compare all approaches on three real-world datasets. Finally, Section~\ref{sec:rw} lists the related work and we conclude in Section~\ref{sec:conclusions}.

%% file: 03_model.tex

\section{Models}%
\label{sec:models}

	In this paper we consider the most intuitive ways to include time-awareness and time modeling in factorization models. First, Section~\ref{sec:baselines} describes the factorization-style models we base our time-aware extensions on. Secondly, Section~\ref{sec:ta} introduces two methods to adapt models to the most recent trends at prediction time. Thirdly, Section~\ref{sec:contributions} presents new models we propose to actually model the flow of time, rather than only adapting to local trends.


	\subsection{Baselines}\label{sec:baselines}

	\subsubsection{Weighted Matrix Factorization --- WMF}
		A popular and very scalable algorithm for modeling implicit feedback data, is known as iALS, or Weighted Matrix Factorization~(WMF)~\cite{hu2008collaborative, pan2008one}. It consists of a simple square loss with a weighting scheme to adjust the importance of each interaction, which helps balance out the abundance of unknowns vs.\ a small portion of positive interactions:
		\begin{equation}
			\lossl{WMF} = \sum_{\substack{u, i \\ X_{ui} = 1}} W_{ui} \left(1 - \Xpred_{ui}\right)^2 + \sum_{\substack{u, i \\ X_{ui} = 0}} {\Xpred_{ui}}^2 + \lambda \mathcal{R}
			\label{eq:sqloss2d}
		\end{equation}
		with $\Xpred_{ui}$ the prediction for user $u$ and item $i$ and $\mathcal{R}$ the regularization term with strength $\lambda$. In the case of WMF, predictions are made by the dot product between user and item latent factors, $\Xpred_{ui}= P_u  Q_i^\top$, and an $\ell^2$ penalty is used for regularization, $\mathcal{R} = \frobnormsq{P} + \frobnormsq{Q}$.
		The weight $W_{ui}$ adjusts the importance of each interaction. Typically a uniform weighting scheme such as $W_{ui} = \alpha X_{ui}$ is used, which already achieves competitive results~\cite{rendle2022revisiting}.
		WMF can be optimized efficiently, and on all data without sampling, through alternating least squares~(ALS), where alternatingly the user factors are fixed and the item factors learned, and vice versa.

		It has also been shown that the loss can be simplified by replacing the second sum, over the unknown interactions, by a sum over all user-item combinations~\cite{bayer2017generic}. Doing so, only results in a scaling of the optimal hyperparameter $\alpha$ and $\lambda$ that would be discovered. In this sense, it acts as a form of implicit regularizer that pulls all predictions towards zero. We continue with this simplification for the next models.

	\subsubsection{Weighted Tensor Factorization --- iTALS and iTALSx}
		The most straightforward way to include the time dimension into a factorization model, is to factorize a three-dimensional tensor, where the third dimension represents time. An illustration of this tensor is given on the left side of Figure~\ref{fig:dtf}. Extending the weighted square loss function of Equation~\eqref{eq:sqloss2d} to include a time dimension results in:
		\begin{equation}
			\loss = \sum_{\substack{u, i, t \\ X_{uit} = 1}} W_{uit} \left(1 - \Xpred_{uit}\right)^2 + \sum_u^m \sum_i^n \frac{1}{l} \sum_t^l {\Xpred_{uit}}^2 + \lambda \mathcal{R} \label{eq:sqloss3d}
		\end{equation}
		Of course, learning a latent representation for each unique point in time is both infeasible, and counterintuitive, because we want the model to generalize to unseen time points. A binning approach is employed to divide the time dimension into $l$ bins of fixed length, one day for example. This way, the model learns a single latent representation for each day, which is then used to predict the interactions for all users and items on that day. To facilitate the hyperparameter search, we scale the implicit regularizer with $\sfrac{1}{l}$. This brings the contribution of this term to the same magnitude as in Equation~\eqref{eq:sqloss2d}.

		Notice however that, unlike with matrix factorization, there are many ways to factorize a tensor into a latent space. An overview of known tensor based methods for context-aware recommendation is given in~\cite{depauw2025wtf, frolov2017tensor}. We study the two simplest and most scalable approaches: iTALS~\cite{hidasi2012fast} and iTALSx~\cite{hidasi2014factorization}. For these models respectively, the predictions are given by:
		\begin{equation*}
			\Xpred^{\textrm{(iTALS)}}_{uit} = \left(P_u \odot Q_{i} \odot B_{t}\right) \vec{1} \qquad\qquad
			\Xpred^{\textrm{(iTALSx)}}_{uit} = P_u Q_i^\top + P_u B_t^\top + Q_i B_t^\top \\
		\end{equation*}
		with $B_t$ the latent representation for time bin $t$ and $\odot$ the element-wise product. For the regularization of iTALS, we use $\normsq{\vec{p_d} - P_u}_2$ with $\vec{p_d}$ the ``default factor'' for user embeddings, which can either be $\vec{0}$ or $\vec{1}$. As explained in~\cite{depauw2025wtf}, better performance can often be achieved by regularizing some factors towards the identity operation (the element-wise product with a vector of ones in this case). For each of the three dimensions independently, a hyperparameter decides which regularization to use. This allows the model to learn that, for example, the user factors are less important, and predictions should mainly be based on the item and time factors.

	\subsubsection{EASE}
		EASE is a full-rank autoencoder with closed-form solution~\cite{steck2019embarrassingly}. We include it in our analysis of factorization methods because it can also be seen as one. Consider its loss function:
		\begin{equation}
			\lossl{EASE} = \frobnormsq{X - XB} + \lambda \frobnormsq{B} \qquad \textrm{s.t.} \diag(B) = 0 \label{eq:ease}
		\end{equation}
		If we assume $X$ to be the fixed and full rank user embeddings, then $B$ serves as item embeddings in a full-rank matrix factorization. However, there are limitations to this comparison: the user embeddings are not learned and they are sparse and binary. This leads to notable differences in, for example, the required regularization strength, and the need for weighting to balance the importance of the positive and negative interactions. Previous studies showed that, counterintuitively, an optimal weight $\alpha=1$ is found when adding weights to EASE~\cite{de2024role, ayoub2025does}, making the weights obsolete.

	\subsubsection{Trending (Popularity-based Recommendations)}\label{sec:trending}
		As a trivial baseline, we also compute the most popular items in the training data and recommend them to all users. A cutoff point in time is used as a hyperparameter to determine the most recently relevant items, rather than simply using the global counts.

	\subsection{Time-aware Adaptations}\label{sec:ta}
		The previously described methods all either ignore time, or treat all interactions over time equally. In practice however, the most recent data is found to be a better proxy of the current user preferences. Two time-aware adaptation methods are described below.

		\subsubsection{Popularity Scaling}\label{sec:popularity_scaling}
			A first approach is based on the assumption that the models have captured global item popularities and we want to adapt them to the most recent trends. Similar to the \emph{Trending} recommender, we compute both the global and local popularities of each item (as counts of interactions), and in a postprocessing step we rescale the predictions by the ratio of the local popularity to the global popularity with exponent $\nu$:
			\begin{equation*}
				\Xpred^\prime_{uit} = \Xpred_{uit} \left(\frac{\textrm{local\_pop}_i}{\textrm{global\_pop}_i}\right)^{\nu}
			\end{equation*}
			Intuitively, this allows models to be trained on the full dataset, where they can capture complex dependencies that require a lot of data to model, while also being adapted to the most recent popularity trends, which can be estimated on much fewer data points~\cite{steck2019collaborative}. 

		\subsubsection{Decay Weighted Popularity}\label{sec:popularity_decay}
			Instead of the uniform weighting scheme, we can also use a decaying weighting scheme, where the importance of each interaction decreases with time. We employ a simple exponential decay function, where the weight of an interaction is halved every \emph{half\_life} time units:
			\begin{equation*}
				W_{uit} = \alpha \cdot 0.5^\frac{t_{\textrm{max}} - t}{\textrm{half\_life}}
			\end{equation*}

			The popularity scaling method is fully model-agnostic and can be used in combination with any model, whereas the decaying weights can only used for weighted loss functions (not EASE). 

	\subsection{Contributions}\label{sec:contributions}
		In the following sections, we present
		three
		new ways to explicitly model the flow of time, based on factorization. The first two methods, DTF and DTFKernel, are generalizations of iTALS to continuous time, where binning is no longer needed. The third method, DMF, adapts WMF to have item factors that change smoothly over time.

	\subsubsection{Discrete-event Tensor Factorization --- DTF}\label{sec:dtf}
	
		In this model, we learn continuous time factors from discrete-event based data, such as interactions, without binning or making any assumptions about the distribution of the underlying relevance signal over time.
		The key insight that allows for continuous time modeling, is that the learned time factors change smoothly over time~\cite{pagano2016contextual, hidasi2014approximate}.

		Indeed, consider binning interactions per day, and inspecting the learned time factors of two consecutive days. One can expect that the data distribution does not change drastically from one day to the next, and that the learned time factors should therefore also be similar. We would also not want the model to learn vastly different latent representation for each day, as the underlying relevance signal we want to capture does not change rapidly over time either.

		The issue with binning approaches however, is that the granularity of the bins can adversely impact the performance of the model. If the bins are too large, the model will not be able to capture trends in the data. On the other hand, if the bins are too small, there is not enough data in each bin to accurately estimate it, and the result will be overfit and noisy. In the DTF model, time factors are \emph{virtual} and are learned jointly as a polynomial fit over the time dimension. This eliminates the issue of choosing the right bin size altogether, and allows for each time factor to also learn from those around it.
		
		\begin{figure}
			\includegraphics[width=0.7\linewidth]{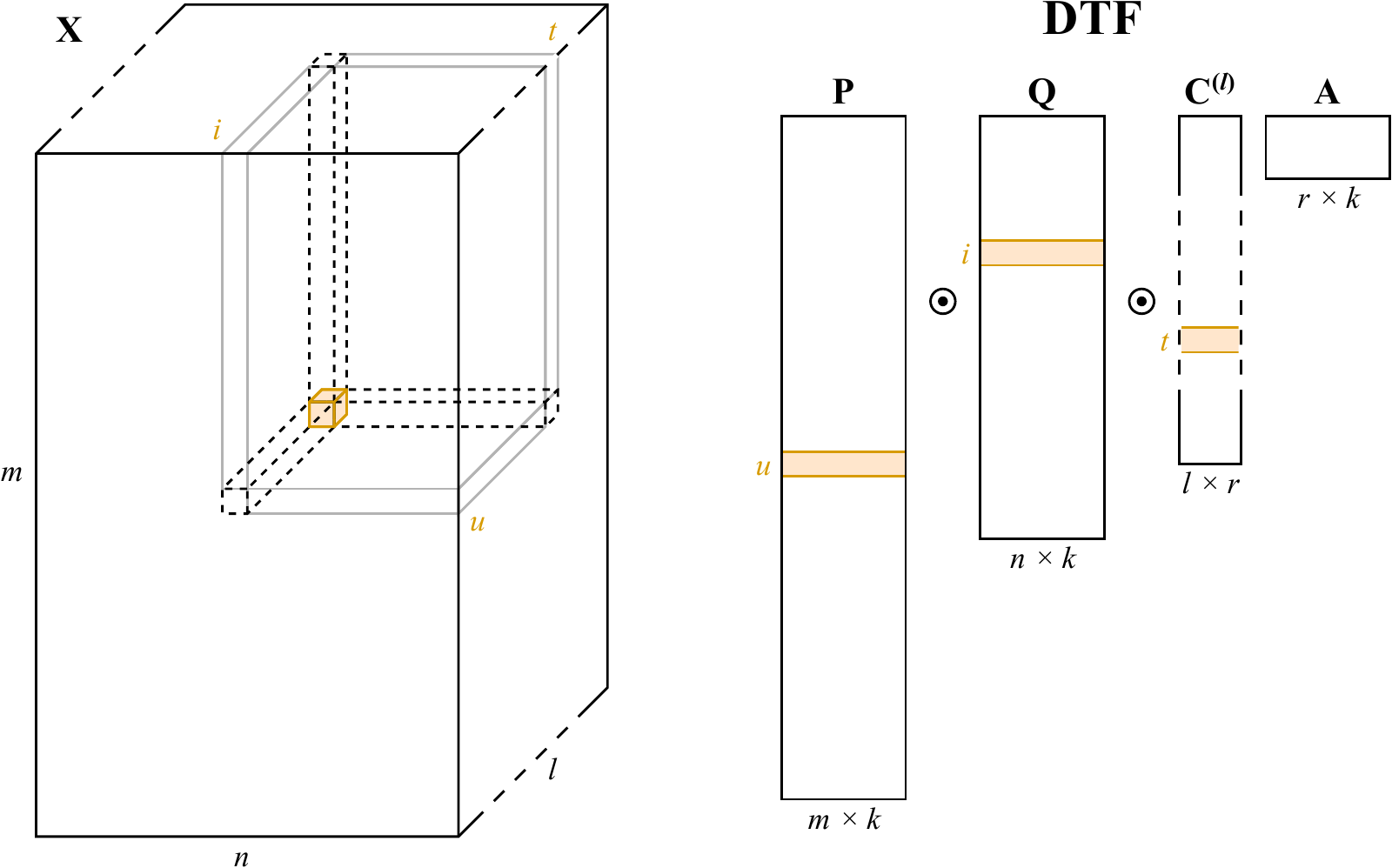}
			\caption{Left: User-item-time interaction tensor with a single entry $X_{uit}$ highlighted. Right: DTF model with a fit over the continuous time dimension through coefficient matrix $A$.}\label{fig:dtf}
			\Description{Visual representation of the DTF model.}
		\end{figure}

		The prediction at timestamp $t$ (not binned) is given by:
		\begin{equation*}
			\Xpred^{\textrm{(DTF)}}_{uit} = \left(P_u \odot Q_i \odot C_t A\right) \vec{1}
		\end{equation*}
		This is based on the iTALS model, with $B_t = C_t A$. Indeed, rather than learning a factor per timestamp, DTF learns a polynomial fit of degree $r-1$ for each dimension of the virtual time factors. The coefficients of these fits, are the weights in matrix $A \in \R^{r \times k}$. Lastly, to perform the least-squares fits, we evaluate the polynomial basis functions at every timestamp, and collect them in the pseudo-Vandermonde matrix $C$.
		Note that $C$ is not learned, as each row is simply an evaluation of the polynomial basis functions of degree $0$ to $r-1$ at a specific timestamp. Then, to obtain the value of the fit at time $t$, all we need is the dot product between $C_t$ and $A$, which results in the time factor of size $k$. A graphical illustration of the model is given in Figure~\ref{fig:dtf}.


		For the polynomial basis functions, we chose the Legendre Polynomials~\cite{weisstein2002legendre}, as they are orthogonal over the interval $[-1, 1]$, which is the interval we normalize the timestamps to. If we denote by $T^{(l)}$ a uniform sample of $l$ timestamps and assume a sum over infinite samples, the loss function of the DTF model is given by:
		\begin{equation*}
			\lossl{DTF} = \sum_{\substack{u, i, t \\ X_{uit} = 1}} W_{uit} \left(1 - \Xpred_{uit}\right)^2 + \sum_u^m \sum_i^n \lim_{l \to \infty} \frac{1}{l} \sum_t^{T^{(l)}} {\Xpred_{uit}}^2 + \mathcal{R}
		\end{equation*}
		The two terms clearly show the learning from discrete-events (positive interactions), and the continuous nature of the learned factors with the limit term. The regularization term is defined as:
		\begin{equation*}
			\mathcal{R} = \lambda \left(\frobnormsq{P} + \frobnormsq{Q} + \lim_{l \to \infty} \frac{1}{l} \frobnormsq{C^{(l)} A}   \right) + \lambda_A \frobnormsq{A}
		\end{equation*}
		where we include additional regularization on the fit coefficients $A$ to prevent overfitting with hyperparameter $\lambda_A$, and denote with $C^{(l)} \in \R^{l \times r}$ the matrix with polynomial basis functions evaluated at the timestamps in $T^{(l)}$. Though the limit terms look impossible to compute in practice, after taking the derivative and setting it to zero, it becomes clear that a simplification is possible.
		
		The update formulas are as follows, with $B_t = C_t A$ to simplify notation, and variables in bold are learned:
		\begin{equation*}
			\hspace{-1em} 
			\biggl((Q^\top Q) \odot (\lim_{l \to \infty} \frac{\Bl^\top \Bl}{l})
				+ \sum_{\substack{i, t \\ X_{uit}=1}} W_{uit} (Q_i^\top Q_i) \odot (B_t^\top B_t)
				+ \lambda I\biggr) \boldsymbol{P_u}^\top 
				= \sum_{\substack{i, t \\ X_{uit}=1}} W_{uit} (Q_i \odot B_t)^\top
		\end{equation*}
		\begin{equation*}
			\hspace{-1em}
			\sum_{\substack{u, i, t \\ X_{uit}=1}} W_{uit} C_t^\top \left(P_u \odot Q_i\right) = \left(\lim_{l \to \infty} \frac{\Cl^\top \Cl}{l} \right) \boldsymbol{A} \left(Q^\top Q \odot P^\top P  + \lambda I \right)
			+ \sum_{\substack{u, i, t \\ X_{uit}=1}} W_{uit}\left(C_t^\top C_t \right) \boldsymbol{A} \left(Q_i^\top Q_i \odot P_u^\top P_u\right)
			+ \lambda_A \boldsymbol{A}
			\hspace{-1em} 
		\end{equation*}
		The formula for $Q_i$ is analogous to $P_u$ and omitted. To solve the limit terms, first note that $\Bl^\top \Bl = A^\top \left(\Cl^\top \Cl\right) A$, and we only need to find the result of:
		\begin{equation*}
			G = \lim_{l \to \infty} \frac{\Cl^\top \Cl}{l} = \int_{-1}^1 C_t^\top C_t \, \diff t
		\end{equation*}
		which is a matrix of size $r \times r$. As shown above, the infinite sum is the definition of a Riemann sum and can be replaced by the integral of pairwise products of the basis functions. Furthermore, due to the orthogonality property of the Legendre polynomials, the integral has a simple analytical solution that is a diagonal matrix with $G_{ii} = \frac{2}{2i + 1}$~\cite{weisstein2002legendre}.
		Lastly, the matrix $A \in \R^{r\times k}$ can either be solved exactly by vectorization~\cite{lancaster1985theory} or with approximate sparse solvers, such as the conjugate gradient method~(CG)~\cite{hestenes1952methods}. Given that $A$ is relatively small (20 by 256 in our experiments), vectorization is actually feasible, but we still adopt the CG approach for its faster and more memory efficient computation.
		The total complexity of one iteration is $\mathcal{O}\left((m + n)  k^3 + p  r^2k^2\right)$, with $p$ the amount of interactions.
		Since DTF does not depend on the amount of bins $l$, it is actually more efficient for large datasets than iTALS.

		\paragraph{Interpretation}
		Putting everything together, DTF learns continuously changing time factors, which are smooth and change gradually over time. The loss function has three parts. First, the discrete-events for positive interactions with weight $W_{uit}$ that model a relevance score of $1$ for positive user, item and time combinations. Secondly, an implicit regularizer that pulls predictions for every user and item combination towards $0$ for \emph{all} timestamps. Thirdly, a regularization term that prevents overfitting. 

		Intuitively, the loss of DTF allows the model to capture exactly when an interaction occurred, which could provide additional information about the interaction. For example, a user might watch the movie `The Matrix' when it was very popular and a large portion of users in the dataset also watched it. Then, there is very little to learn from this interaction, because it was simply popular. If on the other hand, the user watched the movie many years later, this would be a stronger signal that the user is a fan of the movie, and perhaps a bit of a geek.
		The DTF model would capture this information by the fact that all predictions for The Matrix at the time when it is popular are high, meaning that $Q_{\textrm{Matrix}} \odot B_{t_{\textrm{pop}}}$ has a high magnitude, which leads to higher predictions for all users. On the other hand, if the interaction happens at a different time, $Q_{\textrm{Matrix}} \odot B_{t_{\textrm{niche}}}$ will have a lower magnitude and the user factor $P_u$ has to contribute more to achieve a high predicted score. This means the niche interaction had a higher influence on shaping the user factor than the popular interaction.

		A second example is when interactions occur closer together in time, they will also be found more similar by the model. Consider $P_u \odot B_{t_1}$ to be the \emph{effective} user vector at time $t_1$ when the user interacted with item $i$. Then if an interaction with item $j$ occurs at time $t_2$, close to $t_1$, the user vector will not have changed much. In order to achieve a high prediction for both interactions, the model will have to put both $Q_i$ and $Q_j$ close
		to the effective user vector at those times, and they will be more similar to each other. When interactions occur further apart, the effective user factor could have changed to accommodate concept drift or shifting user interests, and the item vectors can be more dissimilar.
		

		\pgfplotstableread[col sep=comma, header=true]{data/Netflix/example_Netflix-DTF2.csv}{\dtffPredictions}
		\pgfplotstableread[col sep=comma, header=true]{data/Netflix/example_Netflix-iTALS_TA2.csv}{\italsPredictions}
		\pgfplotstableread[col sep=comma, header=true]{data/Netflix/example_Netflix-times.csv}{\timestamps}
	
		\begin{figure}
			\pgfplotsset{
				every axis post/.style={
					cycle list name=exotic2,
				},
				Predictions/.style={
					ymin=-0.1, ymax=1,
					xmax={2005-12-31 00:00},
					axis lines = left,
					axis line style=-,
					width=\linewidth,
					height=0.6\linewidth,
					date coordinates in=x,
					ytick={0, 0.5, 1},
					xtick={2000-01-01, 2001-01-01, 2002-01-01, 2003-01-01, 2004-01-01, 2005-01-01, 2006-01-01},
				}
			}
			\centering
			\begin{tikzpicture}
				\begin{axis}[
					Predictions,
					width=0.8\linewidth,
					height=0.4\linewidth,
					ymin=-0.22,
					enlarge x limits={value=0.05, upper},
					enlarge y limits={value=0.15, upper},
					ylabel=iTALS,
					legend pos=north west,
					legend style={font=\footnotesize},
					xticklabels=\empty,
				]
					\addlegendimage{empty legend}
					\addplot+[only marks, mark size=0.1pt, mark=*] table [y index=1, x=time]{\italsPredictions};
					\addplot+[only marks, mark size=0.1pt, mark=*] table [y index=2, x=time]{\italsPredictions};
					\addplot+[only marks, mark size=0.1pt, mark=*] table [y index=3, x=time]{\italsPredictions};
					\addplot+[only marks, mark size=0.1pt, mark=*] table [y index=4, x=time]{\italsPredictions};
					\addplot+[only marks, mark size=0.1pt, mark=*] table [y index=5, x=time]{\italsPredictions};
					\pgfplotsset{cycle list shift=-5}
					\foreach \i in {0,...,4} {
						\addplot+[
							y filter/.expression={\coordindex==\i ? y : nan},
						] table [y expr=-0.1, x=time]{\timestamps};
					}
					\legend{\hspace{-.55cm}\textbf{Items},,,,,,A, B, C, D, E};
				\end{axis}
			\end{tikzpicture}\\
			\begin{tikzpicture}
				\begin{axis}[
					Predictions,
					width=0.8\linewidth,
					height=0.4\linewidth,
					ymin=-0.22,
					enlarge x limits={value=0.05, upper},
					enlarge y limits={value=0.15, upper},
					ylabel=DTF,
					xticklabel=\year,
				]
					\addplot+[smooth, mark=none] table [y index=1, x=time]{\dtffPredictions};
					\addplot+[smooth, mark=none] table [y index=2, x=time]{\dtffPredictions};
					\addplot+[smooth, mark=none] table [y index=3, x=time]{\dtffPredictions};
					\addplot+[smooth, mark=none] table [y index=4, x=time]{\dtffPredictions};
					\addplot+[smooth, mark=none] table [y index=5, x=time]{\dtffPredictions};
					\pgfplotsset{cycle list shift=-5}
					\foreach \i in {0,...,4} {
						\addplot+[
							y filter/.expression={\coordindex==\i ? y : nan},
						] table [y expr=-0.1, x=time]{\timestamps};
					}
				\end{axis}
			\end{tikzpicture}
			\caption{Qualitative analysis of the learned time factors of iTALS and DTF. Five interactions were sampled of a single user of the Netflix dataset. Each line shows the predictions made by the respective model over time and the marks indicate the actual time when the user interacted with each item.}
			\Description{Plot of predictions of different models over time for a 5 sampled interactions of a single user.}
			\label{fig:predictions}
		\end{figure}
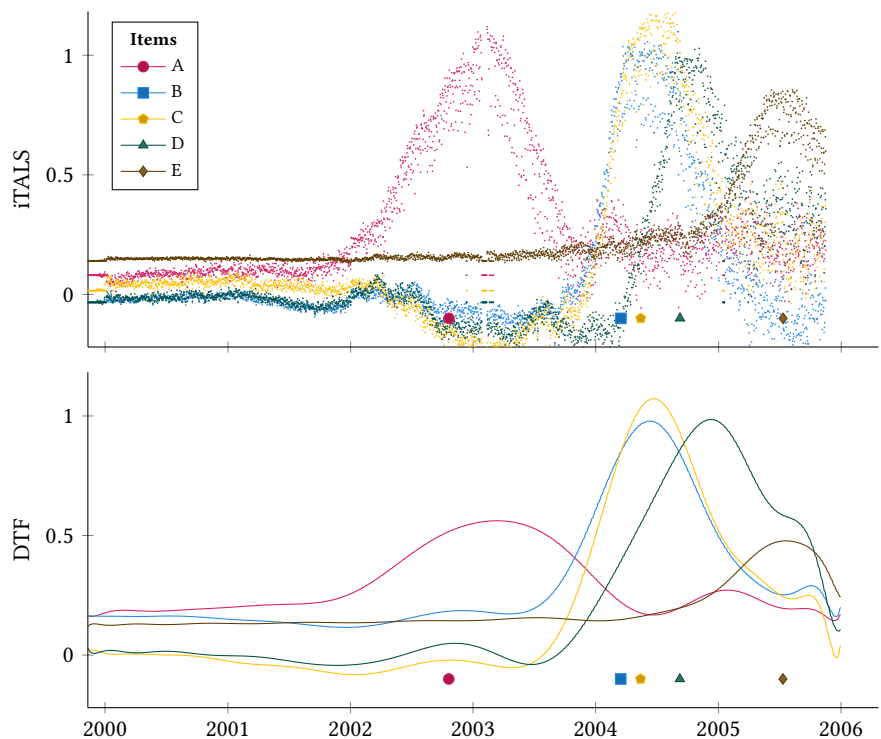

		Figure~\ref{fig:predictions} demonstrates these properties of DTF. It shows how the predictions of each item vary over time, indicating a peak of relevance that was captured around each time an interaction occurred. At the same time, we also see that these are not perfect and independent peaks. Since the ultimate goal of the model is to predict the unknown user-item-time combinations, we actually want the model to generalize across the item and time dimensions. The low-rank approximation makes sure that similar items are modeled close together, and also that a smooth relevance function over time can be estimated from the discrete-events of interactions.
		
		Furthermore, by not explicitly making an assumption about the distribution of the relevance over time, DTF is able to learn from the data itself. This can be seen for example in the difference of heights, widths and shapes of the relevance functions. However, when there is not enough data or when domain knowledge is available about the distribution of the relevance, a kernel function can be used to model the relevance over time. This is described in the next section.

	\subsubsection{DTF with Kernel Function --- DTFKernel}
		An alternative to learning from discrete events, is to make an assumption about the relevance distribution over time. This can be done with a kernel function that models relevance in function of the time since the event. Compared to DTF, this model has less freedom in how it models the relevance over time, but the additional assumption can make it more robust when little data is available. The loss of DTFKernel \emph{only} changes how positive interactions are modeled, as follows:
		\begin{equation*}
			\lossl{DTFKernel} = \sum_{\substack{u, i, t \\ X_{uit} = 1}} W_{uit} \lim_{l \to \infty} \frac{1}{l} \sum_{t^\prime}^{T^{(l)}} \left(\mathcal{K}(t - t^\prime) - \Xpred_{uit}\right)^2
			+ [\ldots]
		\end{equation*}
		Every interaction now fits a value over all timestamps with the sum over $t^\prime$ and the kernel function $\mathcal{K}$ that makes an assumption about the relevance distribution over time. We use the Gaussian kernel: $\exp\left(-(\frac{t - t^\prime}{\sigma})^2\right)$, which models a symmetric peak of relevance around time $t$. The width of the peak is controlled by a hyperparameter $\sigma$.
		Other assumptions can be made, such as a logistic function which approximately models zero relevance before the event and a remaining high relevance after.

		Optimizing this model is very similar to DTF, where only two additional integrals are needed: the integral of the kernel function with specific mean, and the integral of the polynomial basis multiplied by the kernel. Though these integrals can be solved analytically, we found that the numerical integration is less error-prone, and even with 1000 sample points, sufficient accuracy is achieved. Please refer to the source code for more details.

	\subsubsection{Discrete-event Matrix Factorization --- DMF}		
		The idea of learning time factors that vary smoothly over time, can also be applied to, for example, the item factors. Concept drift can be seen as the changing meaning or perception of items over time. Indeed, if we consider $Q_i^t = Q_i \odot B_t$ to be the \emph{effective} item factor at time $t$ in DTF, then rather than model time factors $B_t$, we can also learn a fit for each item directly: $Q_i^t = C_t A_i$. Here, $A_i \in \R^{r \times k}$ is the matrix of coefficients for item $i$, so A is an $n \times r \times k$ tensor of weights that models a fit over time for each item. The loss function of DMF is otherwise identical to that of DTF.

		Optimizing this model is considerably more expensive, as we now have to learn $n$ independent polynomial fits with coefficient matrices. This also means that DMF could be more expressive than DTF with the same embedding dimension, as it has more weights and the item factors can vary more independently of each other to capture their unique behavior. It is even possible to allow the user factors to vary over time as well, however, then the method becomes exceedingly expensive to optimize and the expected benefit is minimal.


	More details about the optimization of every model can be found in the source code: \url{https://github.com/JoeyDP/DTF-Experiments}.

%% file: 05_experiments.tex

\section{Experiments}
\label{sec:experiments}

\begin{table}
    \centering
    \caption{Dataset statistics after preprocessing.}\label{tab:dataset_statistics}
    \begin{tabular}{l lllll}
        \toprule
        \textbf{Dataset} & \textbf{Users} & \textbf{Items} & \textbf{Inter.} & \textbf{Density} & \textbf{Span} \\
        \midrule
        \textbf{Food.com} & \num{22178} & \num{15086} & \num{388362} & \num{0.012}\% & \num{18} yrs \\
		\textbf{Netflix} & \num{475256} & \num{17759} & \num{85723310} & \num{1.02}\% & \num{6} yrs \\
		\textbf{H\&M} & \num{893568} & \num{70792} & \num{27269712} & \num{0.043}\% & \num{2} yrs \\
        \bottomrule
    \end{tabular}
\end{table}

\subsection{Experimental Setup}

Three datasets from various domains and with different characteristics (see Table~\ref{tab:dataset_statistics}) are used in the experiments:
\begin{description}[leftmargin=0cm]
    \item[Food.com~\cite{majumder2019generating}] A website where users rate recipes. Only positive ratings ($\ge3$) were retained and users with fewer than 3 interactions, and items with fewer than 10 interactions, were removed. The validation and test cutoffs are 2009-12-06 and 2011-09-26, which results in \num{8216} and \num{6209} testable users in the validation and test sets.
    \item[Netflix~\cite{netflixdata}] A movie recommendation dataset which is relatively dense (1.02\%). The same preprocessing as for Food.com was applied, and the validation and test cutoffs are 2005-10-02 and 2005-11-16, resulting in \num{215461} and \num{199144} testable users.
    \item[H\&M~\cite{ling2022inversion}] A fashion retail dataset with the highest amount of unique users and items. Users with 5 or more interactions and items with 20 or more interactions were retained. The validation and test cutoffs are 2020-08-23 and 2020-09-07, resulting in \num{132684} and \num{118819} testable users. 
\end{description}

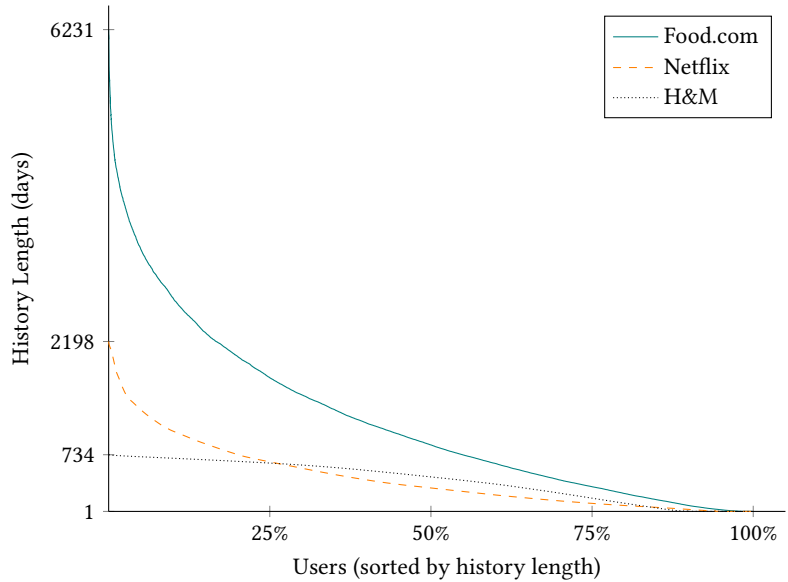
\begin{figure}
    \centering
    \begin{tikzpicture}
        \pgfplotsset{
            ymin=1, ymax=6231,
            every axis post/.style={
                cycle list name=exotic,
                /pgf/number format/.cd,
                1000 sep={},
                enlarge x limits={value=0.05, upper},
                enlarge y limits={value=0.05, upper},
            },
            axis lines = left,
            axis line style=-,
            width=0.7\linewidth,
            height=0.55\linewidth,
        }
        \begin{axis}[
            axis y line=none,
            axis x line=none,
            xmin=0, xmax=22177,
        ]
        \addplot+[smooth, mark=none] table [y={span}, x=index, col sep=comma]{data/Food_com/session_spans.csv}; \label{plt:food}
        \end{axis}

        \begin{axis}[
            xlabel=Users (sorted by history length),
            xticklabel=\empty,
            scaled x ticks = false,
            xtick={118813, 237627, 356441, 475255},
            xticklabels={25\%, 50\%, 75\%, 100\%},
            ylabel=History Length (days),
            ytick={1,734, 2198,6231},
            xmin=0, xmax=475255,    
        ]
        \pgfplotsset{cycle list shift=1}
        \addplot+[smooth, mark=none, dashed] table [y={span}, x=index, col sep=comma]{data/Netflix/session_spans.csv}; \label{plt:netflix}
        \end{axis}
        
        \begin{axis}[
        axis y line=none,
        axis x line=none,
        xmin=0, xmax=893567,
        ]
        \addlegendimage{/pgfplots/refstyle=plt:food}\addlegendentry{Food.com}
        \addlegendimage{/pgfplots/refstyle=plt:netflix}\addlegendentry{Netflix}
        \pgfplotsset{cycle list shift=7}
        \addplot+[smooth, mark=none, densely dotted] table [y={span}, x=index, col sep=comma]{data/H_M/session_spans.csv}; \label{plt:h_m}
        \addlegendentry{H\&M}
        \end{axis}
    \end{tikzpicture}
    \caption{User history length distributions for each dataset. Users are sorted by the length of their history in descending order. The X-axis is relative to account for different numbers of users in the datasets.}\label{fig:session_spans}
    \Description{Long-tail plot of user history lengths. It shows how quickly the history length drops off.}
\end{figure}

These datasets are representative of scenarios where time plays an important role, such as long user histories and varying item lifetimes. To illustrate this, we show the distribution of user history lengths in Figure~\ref{fig:session_spans}. For each user, we take the amount of days between their first and last interactions, and sort them in descending order. In this plot, we can see that it is not uncommon for users to have a history of several years. In all three datasets, more than 40\% of users have a history length of a year or more even, which is representative of real-world scenarios.

To prevent data leakage from the future, we employ a \emph{Temporal Global} split, with global validation and test cutoff points~\cite{ludewig2018evaluation, meng2020exploring}. First, the optimal hyperparameter combinations are determined through Bayesian search~\cite{falkner2018bohb} on the validation set. Optimal parameters and sweep ranges can be found in the source code. Models are trained on all data \emph{before} the validation cutoff, and evaluated per user on their first interaction after the validation cutoff. Secondly, the models are retrained on all data before the test cutoff, and the metrics are reported on the test set, using the same leave-one-out with global time split procedure. This allows us to test whether models generalize to the future, which is the most realistic scenario.

\pgfplotstableread[col sep=comma, header=true]{data/Netflix/metrics.csv}{\netflixMetrics}
\pgfplotstablesort[sort cmp={float <},sort key=NDCG@50]{\netflixMetrics}{\netflixMetrics}
\pgfplotstableread[col sep=comma, header=true]{data/H_M/metrics.csv}{\hmMetrics}
\pgfplotstablesort[sort cmp={float <},sort key=NDCG@50]{\hmMetrics}{\hmMetrics}
\pgfplotstableread[col sep=comma, header=true]{data/Food_com/metrics.csv}{\foodMetrics}
\pgfplotstablesort[sort cmp={float <},sort key=NDCG@50]{\foodMetrics}{\foodMetrics}

\pgfplotsset{
    Metrics/.style={
        xbar,
        set layers,
        clip bounding box=upper bound,
        axis on top,
        bar width=.28cm,
        bar shift=0pt,
        width=0.35\textwidth,
        xmin = 0,
        tick style={draw=none},
        axis line style=-,
        xmajorticks=false,
        xminorticks=false,
        enlarge y limits={abs=0.75},
        enlarge x limits={value=0.18,upper},
        nodes near coords,
        nodes near coords align=right,
        nodes near coords style={
            font=\footnotesize\bfseries,
            },
        ytick=data,
        scaled ticks=false,
        y label style={font=\bfseries},
        x label style={font=\small},
        yticklabel style={anchor=west, font=\footnotesize, inner sep=5pt},
        error bars/x dir=both,
        error bars/x explicit,
        error bars/error bar style={
            line width=0.5pt,
            color=gray,
        },
    },
}

\newcommand*{\datasetBarPlots}[2]{
    \metricBarPlot{#1}{NDCG@50}{#2}
    \metricBarPlot{#1}{MRR@20}{}
    \metricBarPlot{#1}{Recall@20}{}
}

\newcommand*{\metricBarPlot}[3]{
    \pgfplotstablegetrowsof{#1}
    \pgfmathsetmacro{\N}{\pgfplotsretval-1}
    \pgfmathsetmacro{\H}{0.47 * \N}
    \begin{tikzpicture}
        \begin{axis}[
            Metrics,
            yticklabels from table={#1}{Name},
            height=\H cm,
            ytick={0,...,\N},
            xlabel={#2},
            ylabel={#3\strut},
        ]
        \foreach \i in {0,...,\N} {
            \pgfplotstablegetelem{\i}{algorithm}\of{#1}
            \edef\temp{
            \noexpand\addplot [
                fill=\pgfplotsretval,
                y filter/.expression={y==\i ? y : nan},
                fill opacity=0.4,
                text opacity=1,
            ] table [
                y expr=\noexpand\coordindex,
                x={#2},
                x error=#2_err,
                ]{\noexpand#1};
            }
            \temp
        }
        \end{axis}
    \end{tikzpicture}
}

\begin{figure*}
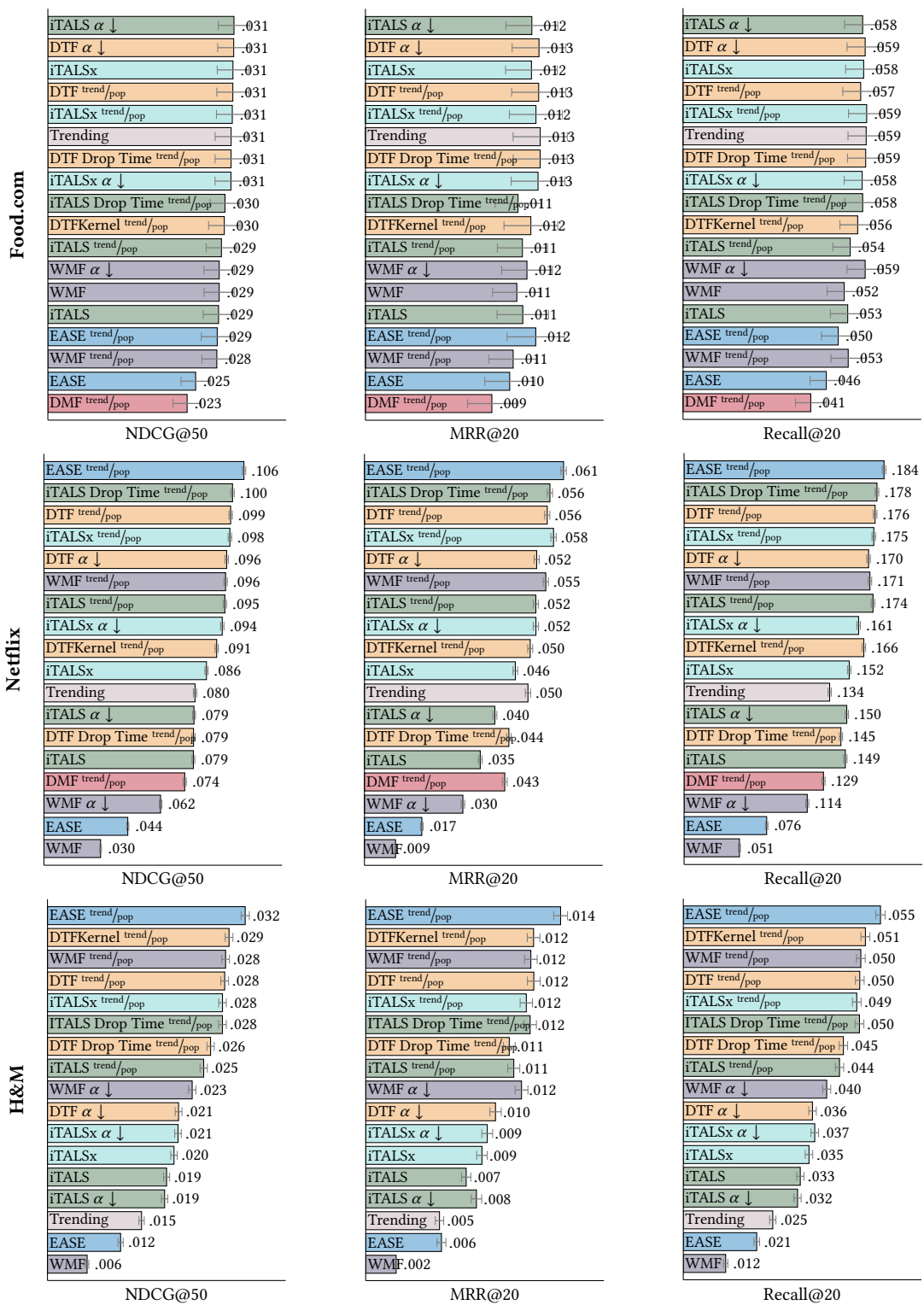

    \pgfkeys{
        /pgf/number format/fixed,
        /pgf/number format/fixed zerofill,
        /pgf/number format/precision=3,
        /pgf/number format/skip 0.,
        /pgf/number format/1000 sep={},
    }
    \centering
    \datasetBarPlots{\foodMetrics}{Food.com}\\
    \datasetBarPlots{\netflixMetrics}{Netflix}\\
    \datasetBarPlots{\hmMetrics}{H\&M}\\
    \caption{Empirical results on three datasets sorted on NDCG@50. Error bars indicate the 95\% confidence intervals. Colors represent different families of algorithms, $\sfrac{\textrm{trend}}{\textrm{pop}}$ refers to popularity scaling~(Section~\ref{sec:popularity_scaling}), and $\alpha\downarrow$ is decay weighting~(Section~\ref{sec:popularity_decay}).}
    \Description{Experimental results}
    \label{fig:metrics}
\end{figure*}

Three well-known metrics are reported: NDCG@50, MRR@20, and Recall@20. Recall and NDCG are computed as described in~\cite{liang2018variational}. MRR is the reciprocal rank of the held-out item in the top-k recommendations, if present. NDCG@50 was the target for the hyperparameter search and a maximum of 10 iterations were used for the optimization, which showed to be sufficient for convergence. Results are reported in Figure~\ref{fig:metrics} with 95\% confidence intervals based on the Student's t-distribution on the individual user scores.

\subsection{Results and Discussion}

    First we consider global trends in the datasets. Food.com is a very small and sparse dataset, which at the same time spans 18 years, making it very hard to provide accurate recommendations. The relatively low number of test users compared to the other two large datasets, leads to a higher uncertainty in the results. In our experiments, no personalized recommendation method was able to surpass the \emph{Trending} baseline, and the results are very close together. We conclude from this that there is no clear signal in this dataset that can be exploited by the models. Hence we cannot draw any conclusions about the time-aware models from this dataset, except for that they do not harm the performance.

    On Netflix and the H\&M datasets the results are more conclusive. The confidence intervals are tighter due to the higher number of test users, which makes the differences between methods statistically significant. EASE $\sfrac{\textrm{trend}}{\textrm{pop}}$ is the best performing model on both datasets. This is in line with previous studies, where it was also found that simple, well-tuned baselines outperform more complex models~\cite{ludewig2018evaluation, steck2019collaborative, ferrari2019we, rendle2022revisiting}.

    However, to the best of our knowledge, no previous work has demonstrated the performance of EASE on a global temporal split. We show that the inclusion of a popularity adjustment to EASE, gives it more than a $+200\%$ increase in metric values, bringing it from the second-to-last position, to the best performing model. It is even able to outperform all our methods that explicitly model time. This indicates that local popularity is the leading-order effect required to compute accurate recommendations.
    Furthermore, as this simple baseline already outperform all the, more sophisticated, time modeling techniques, there is no benefit in comparing our contributions with other state-of-the art temporal models.
    In the next sections we discuss the results of our contributions in more detail.

\subsubsection{Time-aware Adaptations}

    For every type of model, we added variants with \emph{Popularity Scaling} ($\sfrac{\textrm{trend}}{\textrm{pop}}$) and \emph{Decay Weighting} ($\alpha\downarrow$), as described in Section~\ref{sec:ta}.
    Adapting to local trends is shown to be very important in our experimental setup.
    For example, the unpersonalized \emph{Trending} baseline outperforms all time-unaware models in the H\&M dataset, and also some time-aware models on the Netflix dataset. Optimal window lengths are even found to be between $1$ and $20$ days long, which covers a relatively small amount of interactions.
    Furthermore, the best model on the Netflix and H\&M datasets is EASE with popularity scaling. This demonstrates that time-unaware personalization combined with a notion of local trends, is a really powerful baseline. WMF shows the same trend of improvement when adding popularity scaling, and this simple post-prediction adjustment also works significantly better than training with decaying weights. Finally, we can see that even models that explicitly model time, such as iTALS(x) and DTF, can still benefit from this additional push towards learning from the most recent data.

\subsubsection{Binning Time Models: iTALS and iTALSx}

    iTALS (green) and iTALSx (cyan) both model time with explicit factors, by first discretizing the time into bins and then learning a factor for each bin. Because the choice of bin length can have a high impact, and it is data-dependent~\cite{koren2009collaborative}, we experimented with lengths of 1, 7 and 30 days as a hyperparameter. In all cases, optimal results with bins of 1 day could be achieved. For these models as well, the relatively simpler popularity scaling worked better than decaying weights of older interactions. Another improvement could be achieved by ignoring the time factors at prediction time completely, as denoted with \emph{Drop Time}. This idea is explained later in Section~\ref{sec:drop_time}.

\subsubsection{Continuous Time Models: DTF and DTFKernel}

    DTF is an extension of the iTALS model where the time factors are not discretized, but modeled as a polynomial fit. This removes the dependence on a bin size, and allows for a continuous representation of time. As can be seen in Figure~\ref{fig:predictions}, DTF is very effective at capturing the time dynamics of the data. Compared to the binned factors of iTALS, the polynomial fit is able to capture the same signal with a lot less noise, which allows it to generalize better.
    The experiments in Figure~\ref{fig:metrics} validate that, indeed, DTF performs equally well or slightly better than iTALS(x) on all datasets.
    
    For DTFKernel, the results are more divided. On the Netflix dataset it performs worse than DTF, but on H\&M it is able to improve slightly on DTF. A possible explanation, is that the H\&M dataset is much sparser than the Netflix dataset, and that the additional assumption of the kernel function helps the model to generalize from fewer data points.

    Concerning the hyperparameters of DTF and DTFKernel, results showed that a higher regularization weight for the time factors was important to compensate for the $\sfrac{1}{l}$ normalization in the regularization term. In comparison with iTALS, where the regularization is counted for every bin, DTF only counts it once, motivating the need for a higher regularization weight. For the degree of the fit, we found that $r=20$ provides sufficient freedom to learn trends over time, and $r=40$ did not increase the performance on these datasets.
    
    We also found that WMF, the matrix factorization method without time factors, achieves similar results to all time models. This leads us to believe that modeling absolute time does not contribute much to predicting recommendations for future timestamps. As an example, on the H\&M dataset, it turns out that the optimal time factors are all close to the vector of ones. This indicates that the best performance could be achieved by \emph{not} learning a time-dependent signal, similar to the WMF baseline. On the Netflix dataset on the other hand, trends over time were clearly modeled, as shown in Figure~\ref{fig:predictions}.



\subsubsection{DMF}
    The DMF approach differs from DTF in that it learns item factors that evolve independently over time. Unfortunately, it was not able to achieve similar results to the other models and it is clear that its complexity is not justified by the performance. On the H\&M dataset, DMF was even too expensive to compute considering the large amount of items.
    We conclude that, though the hypothesis was valid, there is no benefit to learning a fit per item.

\subsubsection{How to Use Time Factors}\label{sec:drop_time}

    There are different ways to use the time factors at prediction time. We compared the following:
    \begin{description}[leftmargin=1em, topsep=0pt]
        \item[Last Time Factor] This is the default setting for all the reported results. With all test interactions happening in the future, we assume that the last time factor, being the most recent, will be the most informed w.r.t.\ the time of the test interactions.
        \item[Last \emph{N} Time Factors] Here, we take the average of the last $N$ time factors, as a way to smooth out the noise and adapt to the trends of the last few days. This however, did not show any improvement in our experiments (not reported).
        \item[Drop Time] In this method, we ignore time altogether at prediction time, by taking the average over \emph{all} time factors. In combination with the popularity adjustment, this was able to slightly improve the results of iTALS. For DTF, the results decrease. A possible explanation for this, is that the smoothing of the polynomial fit allows DTF to learn a better representation for the last time factor than that of the last bin in iTALS.
        \item[Extrapolate Fit] When a fit is performed, we could also use the actual prediction timestamps, which are in the future. However, given that the test sets only span a short period compared to the training time, the difference is negligible in practice. Additionally, since only zero values are used to train the future timestamps, the predictions will just decrease monotonically compared to the last time factor.
    \end{description}

\subsubsection{Impact of Embedding Dimension}

    All results are reported with an embedding dimension of $k=256$. Due to computational constraints, we were not able to run the full hyperparameter search for all models with a higher embedding size. Still, we did run a few combinations with different embedding dimensions to estimate the trends. For Food.com, we found that even with $k=64$ the same performance could be achieved, indicating that not much can be learned from this dataset.

    On Netflix and H\&M, we found that scaling up to $k=512$ and $k=1024$ could further increase the performance of all factorization models. It is hence possible that they eventually surpass the full-rank EASE model, given a high enough embedding dimension. Nevertheless, the scaling tests did not indicate that the time models benefit more from a higher embedding dimension than the baseline WMF model, so the relative order would be maintained.






%% file: 98_related_work.tex

\section{Related Work}
\label{sec:rw}

``Time'' is a well studied topic in recommender systems, with many sub-fields tackling different aspects of it, such as time-aware RS, streaming models, and dynamic RS~\cite{campos2014time, vinagre2015overview, rana2015study, rabiu2020recommender, al2017review, bogina2023considering}. A recent survey by \citeauthorplain{rabiu2020recommender} first gives an overview of all the roles time can play in RS, and then categorizes models based on how they incorporate the time dimension~\cite{rabiu2020recommender}. In this paper we focused on modeling \emph{absolute} time in matrix factorization based recommender systems, which falls under \emph{time-dependent factor models}. However, most of the reported methods in this category consider relative time or seasonality instead, and none of the related works allow for a fully continuous fit of time. We highlight three of the most relevant works.

First, \citeauthorplain{koren2009collaborative} introduced the first time-aware matrix factorization model, which uses time-dependent bias terms to model the popularity of items, and rating behavior of users over time~\cite{koren2009collaborative}. Secondly, \citeauthorplain{hidasi2014approximate} proposed a method to smoothen learned time factors by either replicating interactions to neighboring bins or to predict with a linear interpolation between bins~\cite{hidasi2014approximate}. They evaluate this in the context of seasonal effects, such as day of week or hour of day. Thirdly, \citeauthorplain{lo2018temporalmf} propose a factorization model where user factors change over time in the context of rating prediction~\cite{lo2018temporalmf}. This is similar to our DMF model where item factors evolve over time, with a different approach for how they transition.

The idea of using basis functions for a smooth decomposition originates from the matrix/tensor interpolation problem~\cite{imaizumi2017tensor, yokota2015smooth}. Here, a dense and small tensor is assumed with uniformly sampled values and few missing entries. By interpolating across one or more smooth axes, the missing values can be approximated more accurately based on the surrounding values. DTF on the other hand, additionally solves the extreme sparsity and non-uniform timestamps porblems that are inherent to recommender systems.

Two other closely related fields are \emph{session-based} recommendation and \emph{sequential recommendation}. Both consider time, either to group interactions into sessions or to order them in a sequence~\cite{wang2021survey, wang2022sequential}. However, the exact time when an interaction occurs, is typically not considered.
A survey by \citeauthorplain{ludewig2018evaluation} found that simple and computationally cheap methods, can often outperform more complex sequential recommendation methods~\cite{ludewig2018evaluation}. A similar trend is observed in our results, where the simple popularity adjustment outperforms more complex approaches that explicitly model time.

SLIST~\cite{choi2021session} and TALE~\cite{park2024temporal} are two recent session-based recommendation methods based on the EASE~\cite{steck2019embarrassingly} model. They adapt the interaction matrix to a session-based format by splitting it in source and target items. A decay method between interaction times encodes relative time and TALE furthermore adds trend-aware normalization which is similar to the popularity adjustment of Section~\ref{sec:popularity_scaling}. Both methods are designed for sequential data, which is not the case in our datasets.





%% file: 99_conclusions.tex

\section{Conclusions}
\label{sec:conclusions}

This paper studied how to encode absolute time in factorization-style recommendation models.
In addition to binning approaches, we also proposed a novel, fully continuous time encoding through a polynomial fit inside the loss function. Our empirical results validate the smooth fit as an alternative to binning, with better generalization and a solution to the sparsity/resolution tradeoff. 
Furthmore, the results show that, simply enforcing local popularity trends is the leading-order effect that can be extracted from time. We conclude from this that matrix factorization is very effective at modeling the seen data, but needs specialized mechanisms for extrapolation to future timestamps.

For future work, we consider modeling seasonality instead (i.e.\ day of week, time of day, etc.). An additional benefit of the smooth fit over binning approaches in this domain, is that periodic basis functions can be used to ensure continuity between the first and last factors.
Furthermore, our proposed methods can potentially be adapted to other applications such as anomaly detection and time series classification.
For example, DTF with its smooth interpolation may find more value when learning from fewer data points, but when a fine-grained time model is still needed.
Lastly, the continuous embedding can also trivially be adapted to deep learning with gradient descent. The weight matrix is simply a dense layer applied after constructing the pseudo-Vandermonde matrix with the timestamps.


%% file: DTF.bbl

\begin{thebibliography}{44}


\ifx \showCODEN    \undefined \def \showCODEN     #1{\unskip}     \fi
\ifx \showDOI      \undefined \def \showDOI       #1{#1}\fi
\ifx \showISBNx    \undefined \def \showISBNx     #1{\unskip}     \fi
\ifx \showISBNxiii \undefined \def \showISBNxiii  #1{\unskip}     \fi
\ifx \showISSN     \undefined \def \showISSN      #1{\unskip}     \fi
\ifx \showLCCN     \undefined \def \showLCCN      #1{\unskip}     \fi
\ifx \shownote     \undefined \def \shownote      #1{#1}          \fi
\ifx \showarticletitle \undefined \def \showarticletitle #1{#1}   \fi
\ifx \showURL      \undefined \def \showURL       {\relax}        \fi
\providecommand\bibfield[2]{#2}
\providecommand\bibinfo[2]{#2}
\providecommand\natexlab[1]{#1}
\providecommand\showeprint[2][]{arXiv:#2}

\bibitem[\protect\citeauthoryear{Al-Hadi, Sharef, Sulaiman, and Mustapha}{Al-Hadi et~al\mbox{.}}{2017}]%
        {al2017review}
\bibfield{author}{\bibinfo{person}{IAAQ Al-Hadi}, \bibinfo{person}{Nurfadhlina~Mohd Sharef}, \bibinfo{person}{Md~Nasir Sulaiman}, {and} \bibinfo{person}{Norwati Mustapha}.} \bibinfo{year}{2017}\natexlab{}.
\newblock \showarticletitle{Review of the temporal recommendation system with matrix factorization}.
\newblock \bibinfo{journal}{\emph{Int. J. Innov. Comput. Inf. Control}} \bibinfo{volume}{13}, \bibinfo{number}{5} (\bibinfo{year}{2017}), \bibinfo{pages}{1579--1594}.
\newblock


\bibitem[\protect\citeauthoryear{Ayoub, Robertson, Liang, Steck, and Kallus}{Ayoub et~al\mbox{.}}{2025}]%
        {ayoub2025does}
\bibfield{author}{\bibinfo{person}{Alex Ayoub}, \bibinfo{person}{Samuel Robertson}, \bibinfo{person}{Dawen Liang}, \bibinfo{person}{Harald Steck}, {and} \bibinfo{person}{Nathan Kallus}.} \bibinfo{year}{2025}\natexlab{}.
\newblock \showarticletitle{Does weighting improve matrix factorization for recommender systems?}. In \bibinfo{booktitle}{\emph{Proceedings of the ACM on Web Conference 2025}}. \bibinfo{pages}{3885--3895}.
\newblock


\bibitem[\protect\citeauthoryear{Bayer, He, Kanagal, and Rendle}{Bayer et~al\mbox{.}}{2017}]%
        {bayer2017generic}
\bibfield{author}{\bibinfo{person}{Immanuel Bayer}, \bibinfo{person}{Xiangnan He}, \bibinfo{person}{Bhargav Kanagal}, {and} \bibinfo{person}{Steffen Rendle}.} \bibinfo{year}{2017}\natexlab{}.
\newblock \showarticletitle{A generic coordinate descent framework for learning from implicit feedback}. In \bibinfo{booktitle}{\emph{Proceedings of the 26th international conference on world wide web}}. \bibinfo{pages}{1341--1350}.
\newblock


\bibitem[\protect\citeauthoryear{Bennett, Elkan, Liu, Smyth, and Tikk}{Bennett et~al\mbox{.}}{2007}]%
        {netflixdata}
\bibfield{author}{\bibinfo{person}{James Bennett}, \bibinfo{person}{Charles Elkan}, \bibinfo{person}{Bing Liu}, \bibinfo{person}{Padhraic Smyth}, {and} \bibinfo{person}{Domonkos Tikk}.} \bibinfo{year}{2007}\natexlab{}.
\newblock \showarticletitle{KDD Cup and workshop 2007}.
\newblock \bibinfo{journal}{\emph{SIGKDD Explor. Newsl.}} \bibinfo{volume}{9}, \bibinfo{number}{2} (\bibinfo{date}{Dec.} \bibinfo{year}{2007}), \bibinfo{pages}{51–52}.
\newblock
\showISSN{1931-0145}
\urldef\tempurl%
\url{https://doi.org/10.1145/1345448.1345459}
\showDOI{\tempurl}


\bibitem[\protect\citeauthoryear{Bogina, Kuflik, Jannach, Bielikova, Kompan, and Trattner}{Bogina et~al\mbox{.}}{2023}]%
        {bogina2023considering}
\bibfield{author}{\bibinfo{person}{Veronika Bogina}, \bibinfo{person}{Tsvi Kuflik}, \bibinfo{person}{Dietmar Jannach}, \bibinfo{person}{Maria Bielikova}, \bibinfo{person}{Michal Kompan}, {and} \bibinfo{person}{Christoph Trattner}.} \bibinfo{year}{2023}\natexlab{}.
\newblock \showarticletitle{Considering temporal aspects in recommender systems: a survey}.
\newblock \bibinfo{journal}{\emph{User Modeling and User-Adapted Interaction}} \bibinfo{volume}{33}, \bibinfo{number}{1} (\bibinfo{year}{2023}), \bibinfo{pages}{81--119}.
\newblock


\bibitem[\protect\citeauthoryear{Campos, D{\'\i}ez, and Cantador}{Campos et~al\mbox{.}}{2014}]%
        {campos2014time}
\bibfield{author}{\bibinfo{person}{Pedro~G Campos}, \bibinfo{person}{Fernando D{\'\i}ez}, {and} \bibinfo{person}{Iv{\'a}n Cantador}.} \bibinfo{year}{2014}\natexlab{}.
\newblock \showarticletitle{Time-aware recommender systems: a comprehensive survey and analysis of existing evaluation protocols}.
\newblock \bibinfo{journal}{\emph{User Modeling and User-Adapted Interaction}} \bibinfo{volume}{24}, \bibinfo{number}{1} (\bibinfo{year}{2014}), \bibinfo{pages}{67--119}.
\newblock


\bibitem[\protect\citeauthoryear{Choi, Kim, Lee, Shim, and Lee}{Choi et~al\mbox{.}}{2021}]%
        {choi2021session}
\bibfield{author}{\bibinfo{person}{Minjin Choi}, \bibinfo{person}{Jinhong Kim}, \bibinfo{person}{Joonseok Lee}, \bibinfo{person}{Hyunjung Shim}, {and} \bibinfo{person}{Jongwuk Lee}.} \bibinfo{year}{2021}\natexlab{}.
\newblock \showarticletitle{Session-aware linear item-item models for session-based recommendation}. In \bibinfo{booktitle}{\emph{Proceedings of the Web Conference 2021}}. \bibinfo{pages}{2186--2197}.
\newblock


\bibitem[\protect\citeauthoryear{De~Pauw and Goethals}{De~Pauw and Goethals}{2024}]%
        {de2024role}
\bibfield{author}{\bibinfo{person}{Joey De~Pauw} {and} \bibinfo{person}{Bart Goethals}.} \bibinfo{year}{2024}\natexlab{}.
\newblock \showarticletitle{The Role of Unknown Interactions in Implicit Matrix Factorization—A Probabilistic View}. In \bibinfo{booktitle}{\emph{Proceedings of the 18th ACM Conference on Recommender Systems}}. \bibinfo{pages}{219--227}.
\newblock


\bibitem[\protect\citeauthoryear{Ding and Li}{Ding and Li}{2005}]%
        {ding2005time}
\bibfield{author}{\bibinfo{person}{Yi Ding} {and} \bibinfo{person}{Xue Li}.} \bibinfo{year}{2005}\natexlab{}.
\newblock \showarticletitle{Time weight collaborative filtering}. In \bibinfo{booktitle}{\emph{Proceedings of the 14th ACM international conference on Information and knowledge management}}. \bibinfo{pages}{485--492}.
\newblock


\bibitem[\protect\citeauthoryear{Falkner, Klein, and Hutter}{Falkner et~al\mbox{.}}{2018}]%
        {falkner2018bohb}
\bibfield{author}{\bibinfo{person}{Stefan Falkner}, \bibinfo{person}{Aaron Klein}, {and} \bibinfo{person}{Frank Hutter}.} \bibinfo{year}{2018}\natexlab{}.
\newblock \bibinfo{title}{BOHB: Robust and Efficient Hyperparameter Optimization at Scale}.
\newblock
\newblock
\showeprint[arxiv]{1807.01774}~[cs.LG]
\urldef\tempurl%
\url{https://arxiv.org/abs/1807.01774}
\showURL{%
\tempurl}


\bibitem[\protect\citeauthoryear{Ferrari~Dacrema, Cremonesi, and Jannach}{Ferrari~Dacrema et~al\mbox{.}}{2019}]%
        {ferrari2019we}
\bibfield{author}{\bibinfo{person}{Maurizio Ferrari~Dacrema}, \bibinfo{person}{Paolo Cremonesi}, {and} \bibinfo{person}{Dietmar Jannach}.} \bibinfo{year}{2019}\natexlab{}.
\newblock \showarticletitle{Are we really making much progress? A worrying analysis of recent neural recommendation approaches}. In \bibinfo{booktitle}{\emph{Proceedings of the 13th ACM conference on recommender systems}}. \bibinfo{pages}{101--109}.
\newblock


\bibitem[\protect\citeauthoryear{Frolov and Oseledets}{Frolov and Oseledets}{2017}]%
        {frolov2017tensor}
\bibfield{author}{\bibinfo{person}{Evgeny Frolov} {and} \bibinfo{person}{Ivan Oseledets}.} \bibinfo{year}{2017}\natexlab{}.
\newblock \showarticletitle{Tensor methods and recommender systems}.
\newblock \bibinfo{journal}{\emph{Wiley Interdisciplinary Reviews: Data Mining and Knowledge Discovery}} \bibinfo{volume}{7}, \bibinfo{number}{3} (\bibinfo{year}{2017}), \bibinfo{pages}{e1201}.
\newblock


\bibitem[\protect\citeauthoryear{Garg, Gupta, Malhotra, Vig, and Shroff}{Garg et~al\mbox{.}}{2019}]%
        {garg2019sequence}
\bibfield{author}{\bibinfo{person}{Diksha Garg}, \bibinfo{person}{Priyanka Gupta}, \bibinfo{person}{Pankaj Malhotra}, \bibinfo{person}{Lovekesh Vig}, {and} \bibinfo{person}{Gautam Shroff}.} \bibinfo{year}{2019}\natexlab{}.
\newblock \showarticletitle{Sequence and time aware neighborhood for session-based recommendations: STAN}. In \bibinfo{booktitle}{\emph{Proceedings of the 42nd International ACM SIGIR Conference on Research and Development in Information Retrieval}}. \bibinfo{pages}{1069--1072}.
\newblock


\bibitem[\protect\citeauthoryear{Hestenes and Stiefel}{Hestenes and Stiefel}{1952}]%
        {hestenes1952methods}
\bibfield{author}{\bibinfo{person}{Magnus~R Hestenes} {and} \bibinfo{person}{Eduard Stiefel}.} \bibinfo{year}{1952}\natexlab{}.
\newblock \showarticletitle{Methods of conjugate gradients for solving}.
\newblock \bibinfo{journal}{\emph{Journal of research of the National Bureau of Standards}} \bibinfo{volume}{49}, \bibinfo{number}{6} (\bibinfo{year}{1952}), \bibinfo{pages}{409}.
\newblock


\bibitem[\protect\citeauthoryear{Hidasi}{Hidasi}{2014}]%
        {hidasi2014factorization}
\bibfield{author}{\bibinfo{person}{Bal{\'a}zs Hidasi}.} \bibinfo{year}{2014}\natexlab{}.
\newblock \showarticletitle{Factorization models for context-aware recommendations}.
\newblock \bibinfo{journal}{\emph{Infocommun J VI (4)}} (\bibinfo{year}{2014}), \bibinfo{pages}{27--34}.
\newblock


\bibitem[\protect\citeauthoryear{Hidasi and Tikk}{Hidasi and Tikk}{2012}]%
        {hidasi2012fast}
\bibfield{author}{\bibinfo{person}{Bal{\'a}zs Hidasi} {and} \bibinfo{person}{Domonkos Tikk}.} \bibinfo{year}{2012}\natexlab{}.
\newblock \showarticletitle{Fast ALS-based tensor factorization for context-aware recommendation from implicit feedback}. In \bibinfo{booktitle}{\emph{Joint European Conference on Machine Learning and Knowledge Discovery in Databases}}. Springer, \bibinfo{pages}{67--82}.
\newblock


\bibitem[\protect\citeauthoryear{Hidasi and Tikk}{Hidasi and Tikk}{2014}]%
        {hidasi2014approximate}
\bibfield{author}{\bibinfo{person}{Bal{\'a}zs Hidasi} {and} \bibinfo{person}{Domonkos Tikk}.} \bibinfo{year}{2014}\natexlab{}.
\newblock \showarticletitle{Approximate modeling of continuous context in factorization algorithms}. In \bibinfo{booktitle}{\emph{Proceedings of the 4th Workshop on Context-Awareness in Retrieval and Recommendation}}. \bibinfo{pages}{3--9}.
\newblock


\bibitem[\protect\citeauthoryear{Hu, Koren, and Volinsky}{Hu et~al\mbox{.}}{2008}]%
        {hu2008collaborative}
\bibfield{author}{\bibinfo{person}{Yifan Hu}, \bibinfo{person}{Yehuda Koren}, {and} \bibinfo{person}{Chris Volinsky}.} \bibinfo{year}{2008}\natexlab{}.
\newblock \showarticletitle{Collaborative filtering for implicit feedback datasets}. In \bibinfo{booktitle}{\emph{2008 Eighth IEEE international conference on data mining}}. Ieee, \bibinfo{pages}{263--272}.
\newblock


\bibitem[\protect\citeauthoryear{Imaizumi and Hayashi}{Imaizumi and Hayashi}{2017}]%
        {imaizumi2017tensor}
\bibfield{author}{\bibinfo{person}{Masaaki Imaizumi} {and} \bibinfo{person}{Kohei Hayashi}.} \bibinfo{year}{2017}\natexlab{}.
\newblock \showarticletitle{Tensor decomposition with smoothness}. In \bibinfo{booktitle}{\emph{International conference on machine learning}}. PMLR, \bibinfo{pages}{1597--1606}.
\newblock


\bibitem[\protect\citeauthoryear{Koren}{Koren}{2009}]%
        {koren2009collaborative}
\bibfield{author}{\bibinfo{person}{Yehuda Koren}.} \bibinfo{year}{2009}\natexlab{}.
\newblock \showarticletitle{Collaborative filtering with temporal dynamics}. In \bibinfo{booktitle}{\emph{Proceedings of the 15th ACM SIGKDD International Conference on Knowledge Discovery and Data Mining}} (Paris, France) \emph{(\bibinfo{series}{KDD '09})}. \bibinfo{publisher}{Association for Computing Machinery}, \bibinfo{address}{New York, NY, USA}, \bibinfo{pages}{447–456}.
\newblock
\showISBNx{9781605584959}
\urldef\tempurl%
\url{https://doi.org/10.1145/1557019.1557072}
\showDOI{\tempurl}


\bibitem[\protect\citeauthoryear{Lancaster and Tismenetsky}{Lancaster and Tismenetsky}{1985}]%
        {lancaster1985theory}
\bibfield{author}{\bibinfo{person}{Peter Lancaster} {and} \bibinfo{person}{Miron Tismenetsky}.} \bibinfo{year}{1985}\natexlab{}.
\newblock \bibinfo{booktitle}{\emph{The theory of matrices: with applications}}.
\newblock \bibinfo{publisher}{Elsevier}.
\newblock


\bibitem[\protect\citeauthoryear{Liang, Krishnan, Hoffman, and Jebara}{Liang et~al\mbox{.}}{2018}]%
        {liang2018variational}
\bibfield{author}{\bibinfo{person}{Dawen Liang}, \bibinfo{person}{Rahul~G Krishnan}, \bibinfo{person}{Matthew~D Hoffman}, {and} \bibinfo{person}{Tony Jebara}.} \bibinfo{year}{2018}\natexlab{}.
\newblock \showarticletitle{Variational autoencoders for collaborative filtering}. In \bibinfo{booktitle}{\emph{Proceedings of the 2018 world wide web conference}}. \bibinfo{pages}{689--698}.
\newblock


\bibitem[\protect\citeauthoryear{Ling, HMGroup, and Rim}{Ling et~al\mbox{.}}{2022}]%
        {ling2022inversion}
\bibfield{author}{\bibinfo{person}{Carlos~Garc{\'\i}a Ling}, \bibinfo{person}{Elizabeth HMGroup}, {and} \bibinfo{person}{Frida Rim}.} \bibinfo{year}{2022}\natexlab{}.
\newblock \showarticletitle{inversion, Jaime Ferrando, Maggie, neuraloverflow, xlsrln}.
\newblock \bibinfo{journal}{\emph{H\&M Personalized Fashion Recommendations. Kaggle. https://kaggle. com/competitions/h-and-m-personalized-fashion-recommendations}} (\bibinfo{year}{2022}).
\newblock


\bibitem[\protect\citeauthoryear{Lo, Liao, Chang, and Lee}{Lo et~al\mbox{.}}{2018}]%
        {lo2018temporalmf}
\bibfield{author}{\bibinfo{person}{Yung-Yin Lo}, \bibinfo{person}{Wanjiun Liao}, \bibinfo{person}{Cheng-Shang Chang}, {and} \bibinfo{person}{Ying-Chin Lee}.} \bibinfo{year}{2018}\natexlab{}.
\newblock \showarticletitle{Temporal Matrix Factorization for Tracking Concept Drift in Individual User Preferences}.
\newblock \bibinfo{journal}{\emph{IEEE Transactions on Computational Social Systems}} \bibinfo{volume}{5}, \bibinfo{number}{1} (\bibinfo{year}{2018}), \bibinfo{pages}{156--168}.
\newblock
\urldef\tempurl%
\url{https://doi.org/10.1109/TCSS.2017.2772295}
\showDOI{\tempurl}


\bibitem[\protect\citeauthoryear{Ludewig and Jannach}{Ludewig and Jannach}{2018}]%
        {ludewig2018evaluation}
\bibfield{author}{\bibinfo{person}{Malte Ludewig} {and} \bibinfo{person}{Dietmar Jannach}.} \bibinfo{year}{2018}\natexlab{}.
\newblock \showarticletitle{Evaluation of session-based recommendation algorithms}.
\newblock \bibinfo{journal}{\emph{User Modeling and User-Adapted Interaction}}  \bibinfo{volume}{28} (\bibinfo{year}{2018}), \bibinfo{pages}{331--390}.
\newblock


\bibitem[\protect\citeauthoryear{Majumder, Li, Ni, and McAuley}{Majumder et~al\mbox{.}}{2019}]%
        {majumder2019generating}
\bibfield{author}{\bibinfo{person}{Bodhisattwa~Prasad Majumder}, \bibinfo{person}{Shuyang Li}, \bibinfo{person}{Jianmo Ni}, {and} \bibinfo{person}{Julian McAuley}.} \bibinfo{year}{2019}\natexlab{}.
\newblock \showarticletitle{Generating personalized recipes from historical user preferences}.
\newblock \bibinfo{journal}{\emph{arXiv preprint arXiv:1909.00105}} (\bibinfo{year}{2019}).
\newblock


\bibitem[\protect\citeauthoryear{Matuszyk, Vinagre, Spiliopoulou, Jorge, and Gama}{Matuszyk et~al\mbox{.}}{2015}]%
        {matuszyk2015forgetting}
\bibfield{author}{\bibinfo{person}{Pawel Matuszyk}, \bibinfo{person}{Jo{\~a}o Vinagre}, \bibinfo{person}{Myra Spiliopoulou}, \bibinfo{person}{Al{\'\i}pio~M{\'a}rio Jorge}, {and} \bibinfo{person}{Jo{\~a}o Gama}.} \bibinfo{year}{2015}\natexlab{}.
\newblock \showarticletitle{Forgetting methods for incremental matrix factorization in recommender systems}. In \bibinfo{booktitle}{\emph{Proceedings of the 30th annual ACM symposium on applied computing}}. \bibinfo{pages}{947--953}.
\newblock


\bibitem[\protect\citeauthoryear{Meng, McCreadie, Macdonald, and Ounis}{Meng et~al\mbox{.}}{2020}]%
        {meng2020exploring}
\bibfield{author}{\bibinfo{person}{Zaiqiao Meng}, \bibinfo{person}{Richard McCreadie}, \bibinfo{person}{Craig Macdonald}, {and} \bibinfo{person}{Iadh Ounis}.} \bibinfo{year}{2020}\natexlab{}.
\newblock \showarticletitle{Exploring data splitting strategies for the evaluation of recommendation models}. In \bibinfo{booktitle}{\emph{Proceedings of the 14th acm conference on recommender systems}}. \bibinfo{pages}{681--686}.
\newblock


\bibitem[\protect\citeauthoryear{Pagano, Cremonesi, Larson, Hidasi, Tikk, Karatzoglou, and Quadrana}{Pagano et~al\mbox{.}}{2016}]%
        {pagano2016contextual}
\bibfield{author}{\bibinfo{person}{Roberto Pagano}, \bibinfo{person}{Paolo Cremonesi}, \bibinfo{person}{Martha Larson}, \bibinfo{person}{Bal{\'a}zs Hidasi}, \bibinfo{person}{Domonkos Tikk}, \bibinfo{person}{Alexandros Karatzoglou}, {and} \bibinfo{person}{Massimo Quadrana}.} \bibinfo{year}{2016}\natexlab{}.
\newblock \showarticletitle{The contextual turn: From context-aware to context-driven recommender systems}. In \bibinfo{booktitle}{\emph{Proceedings of the 10th ACM conference on recommender systems}}. \bibinfo{pages}{249--252}.
\newblock


\bibitem[\protect\citeauthoryear{Pan, Zhou, Cao, Liu, Lukose, Scholz, and Yang}{Pan et~al\mbox{.}}{2008}]%
        {pan2008one}
\bibfield{author}{\bibinfo{person}{Rong Pan}, \bibinfo{person}{Yunhong Zhou}, \bibinfo{person}{Bin Cao}, \bibinfo{person}{Nathan~N Liu}, \bibinfo{person}{Rajan Lukose}, \bibinfo{person}{Martin Scholz}, {and} \bibinfo{person}{Qiang Yang}.} \bibinfo{year}{2008}\natexlab{}.
\newblock \showarticletitle{One-class collaborative filtering}. In \bibinfo{booktitle}{\emph{2008 Eighth IEEE international conference on data mining}}. IEEE, \bibinfo{pages}{502--511}.
\newblock


\bibitem[\protect\citeauthoryear{Park, Yoon, Choi, and Lee}{Park et~al\mbox{.}}{2024}]%
        {park2024temporal}
\bibfield{author}{\bibinfo{person}{Seongmin Park}, \bibinfo{person}{Mincheol Yoon}, \bibinfo{person}{Minjin Choi}, {and} \bibinfo{person}{Jongwuk Lee}.} \bibinfo{year}{2024}\natexlab{}.
\newblock \showarticletitle{Temporal Linear Item-Item Model for Sequential Recommendation}.
\newblock \bibinfo{journal}{\emph{arXiv preprint arXiv:2412.07382}} (\bibinfo{year}{2024}).
\newblock


\bibitem[\protect\citeauthoryear{Pauw and Goethals}{Pauw and Goethals}{2025}]%
        {depauw2025wtf}
\bibfield{author}{\bibinfo{person}{Joey~De Pauw} {and} \bibinfo{person}{Bart Goethals}.} \bibinfo{year}{2025}\natexlab{}.
\newblock \bibinfo{title}{Weighted Tensor Decompositions for Context-aware Collaborative Filtering}.
\newblock
\newblock
\showeprint[arxiv]{2503.08393}~[cs.IR]
\urldef\tempurl%
\url{https://arxiv.org/abs/2503.08393}
\showURL{%
\tempurl}


\bibitem[\protect\citeauthoryear{Rabiu, Salim, Da’u, and Osman}{Rabiu et~al\mbox{.}}{2020}]%
        {rabiu2020recommender}
\bibfield{author}{\bibinfo{person}{Idris Rabiu}, \bibinfo{person}{Naomie Salim}, \bibinfo{person}{Aminu Da’u}, {and} \bibinfo{person}{Akram Osman}.} \bibinfo{year}{2020}\natexlab{}.
\newblock \showarticletitle{Recommender system based on temporal models: a systematic review}.
\newblock \bibinfo{journal}{\emph{Applied Sciences}} \bibinfo{volume}{10}, \bibinfo{number}{7} (\bibinfo{year}{2020}), \bibinfo{pages}{2204}.
\newblock


\bibitem[\protect\citeauthoryear{Rana and Jain}{Rana and Jain}{2015}]%
        {rana2015study}
\bibfield{author}{\bibinfo{person}{Chhavi Rana} {and} \bibinfo{person}{Sanjay~Kumar Jain}.} \bibinfo{year}{2015}\natexlab{}.
\newblock \showarticletitle{A study of the dynamic features of recommender systems}.
\newblock \bibinfo{journal}{\emph{Artificial Intelligence Review}}  \bibinfo{volume}{43} (\bibinfo{year}{2015}), \bibinfo{pages}{141--153}.
\newblock


\bibitem[\protect\citeauthoryear{Rendle, Krichene, Zhang, and Koren}{Rendle et~al\mbox{.}}{2022}]%
        {rendle2022revisiting}
\bibfield{author}{\bibinfo{person}{Steffen Rendle}, \bibinfo{person}{Walid Krichene}, \bibinfo{person}{Li Zhang}, {and} \bibinfo{person}{Yehuda Koren}.} \bibinfo{year}{2022}\natexlab{}.
\newblock \showarticletitle{Revisiting the performance of ials on item recommendation benchmarks}. In \bibinfo{booktitle}{\emph{Proceedings of the 16th ACM Conference on Recommender Systems}}. \bibinfo{pages}{427--435}.
\newblock


\bibitem[\protect\citeauthoryear{Steck}{Steck}{2019a}]%
        {steck2019collaborative}
\bibfield{author}{\bibinfo{person}{Harald Steck}.} \bibinfo{year}{2019}\natexlab{a}.
\newblock \showarticletitle{Collaborative filtering via high-dimensional regression}.
\newblock \bibinfo{journal}{\emph{arXiv preprint arXiv:1904.13033}} (\bibinfo{year}{2019}).
\newblock


\bibitem[\protect\citeauthoryear{Steck}{Steck}{2019b}]%
        {steck2019embarrassingly}
\bibfield{author}{\bibinfo{person}{Harald Steck}.} \bibinfo{year}{2019}\natexlab{b}.
\newblock \showarticletitle{Embarrassingly shallow autoencoders for sparse data}. In \bibinfo{booktitle}{\emph{The World Wide Web Conference}}. \bibinfo{pages}{3251--3257}.
\newblock


\bibitem[\protect\citeauthoryear{Sun and Dong}{Sun and Dong}{2017}]%
        {sun2017dynamic}
\bibfield{author}{\bibinfo{person}{Baoshan Sun} {and} \bibinfo{person}{Lingyu Dong}.} \bibinfo{year}{2017}\natexlab{}.
\newblock \showarticletitle{Dynamic model adaptive to user interest drift based on cluster and nearest neighbors}.
\newblock \bibinfo{journal}{\emph{IEEE access}}  \bibinfo{volume}{5} (\bibinfo{year}{2017}), \bibinfo{pages}{1682--1691}.
\newblock


\bibitem[\protect\citeauthoryear{Verachtert, Michiels, and Goethals}{Verachtert et~al\mbox{.}}{2022}]%
        {verachtert2022we}
\bibfield{author}{\bibinfo{person}{Robin Verachtert}, \bibinfo{person}{Lien Michiels}, {and} \bibinfo{person}{Bart Goethals}.} \bibinfo{year}{2022}\natexlab{}.
\newblock \showarticletitle{Are We Forgetting Something? Correctly Evaluate a Recommender System With an Optimal Training Window.}. In \bibinfo{booktitle}{\emph{{Perspectives@RecSys}}}.
\newblock


\bibitem[\protect\citeauthoryear{Vinagre, Jorge, and Gama}{Vinagre et~al\mbox{.}}{2015}]%
        {vinagre2015overview}
\bibfield{author}{\bibinfo{person}{Jo{\~a}o Vinagre}, \bibinfo{person}{Al{\'\i}pio~M{\'a}rio Jorge}, {and} \bibinfo{person}{Jo{\~a}o Gama}.} \bibinfo{year}{2015}\natexlab{}.
\newblock \showarticletitle{An overview on the exploitation of time in collaborative filtering}.
\newblock \bibinfo{journal}{\emph{Wiley interdisciplinary reviews: Data mining and knowledge discovery}} \bibinfo{volume}{5}, \bibinfo{number}{5} (\bibinfo{year}{2015}), \bibinfo{pages}{195--215}.
\newblock


\bibitem[\protect\citeauthoryear{Wang, Cao, Wang, Sheng, Orgun, and Lian}{Wang et~al\mbox{.}}{2021}]%
        {wang2021survey}
\bibfield{author}{\bibinfo{person}{Shoujin Wang}, \bibinfo{person}{Longbing Cao}, \bibinfo{person}{Yan Wang}, \bibinfo{person}{Quan~Z Sheng}, \bibinfo{person}{Mehmet~A Orgun}, {and} \bibinfo{person}{Defu Lian}.} \bibinfo{year}{2021}\natexlab{}.
\newblock \showarticletitle{A survey on session-based recommender systems}.
\newblock \bibinfo{journal}{\emph{ACM Computing Surveys (CSUR)}} \bibinfo{volume}{54}, \bibinfo{number}{7} (\bibinfo{year}{2021}), \bibinfo{pages}{1--38}.
\newblock


\bibitem[\protect\citeauthoryear{Wang, Zhang, Hu, Zhang, Wang, and Aggarwal}{Wang et~al\mbox{.}}{2022}]%
        {wang2022sequential}
\bibfield{author}{\bibinfo{person}{Shoujin Wang}, \bibinfo{person}{Qi Zhang}, \bibinfo{person}{Liang Hu}, \bibinfo{person}{Xiuzhen Zhang}, \bibinfo{person}{Yan Wang}, {and} \bibinfo{person}{Charu Aggarwal}.} \bibinfo{year}{2022}\natexlab{}.
\newblock \showarticletitle{Sequential/session-based recommendations: Challenges, approaches, applications and opportunities}. In \bibinfo{booktitle}{\emph{Proceedings of the 45th international ACM SIGIR conference on research and development in information retrieval}}. \bibinfo{pages}{3425--3428}.
\newblock


\bibitem[\protect\citeauthoryear{Weisstein}{Weisstein}{2002}]%
        {weisstein2002legendre}
\bibfield{author}{\bibinfo{person}{Eric~W Weisstein}.} \bibinfo{year}{2002}\natexlab{}.
\newblock \showarticletitle{Legendre polynomial}.
\newblock \bibinfo{journal}{\emph{https://mathworld. wolfram. com/}} (\bibinfo{year}{2002}).
\newblock


\bibitem[\protect\citeauthoryear{Yokota, Zdunek, Cichocki, and Yamashita}{Yokota et~al\mbox{.}}{2015}]%
        {yokota2015smooth}
\bibfield{author}{\bibinfo{person}{Tatsuya Yokota}, \bibinfo{person}{Rafal Zdunek}, \bibinfo{person}{Andrzej Cichocki}, {and} \bibinfo{person}{Yukihiko Yamashita}.} \bibinfo{year}{2015}\natexlab{}.
\newblock \showarticletitle{Smooth nonnegative matrix and tensor factorizations for robust multi-way data analysis}.
\newblock \bibinfo{journal}{\emph{Signal Processing}}  \bibinfo{volume}{113} (\bibinfo{year}{2015}), \bibinfo{pages}{234--249}.
\newblock


\end{thebibliography}
